# Develop a "perfect" spectrometer in "water window"

Zhuo Li[1], Bin Li[1,2*]

[1]*Shanghai Institute of Applied Physics, Chinese Academy of Sciences, Shanghai 201204, China*

[2]*School of Physical Science and Technology, ShanghaiTech University, 100 Haike Road, Shanghai 200031, China*

*Corresponding Author: libin1995@sinap.ac.cn*

We report a novelly universal scheme for design of a grazing incidence spectrometer, which creates an excellent meridional flat field in its detector domain to deliver the desired spectral resolution throughout the full designated spectral range, while eliminates the sagittal astigmatism to enhance the spectral intensity simultaneously. The intrinsic optical properties of the system are thoroughly investigated and optimized, especially how the detector plane approaches to the actual meridional or sagittal focal curves is well quantified. We demonstrated within the "water window" (i.e. 2-5nm wavelength range), the resolving power of 6000-18000 could be achieved for the effective meridional source size of $200 \mu m$ (rms); and it would be further improved to 20000-40000 if the source size is confined down to $50 \mu m$ (rms).

## Ⅰ. Introduction

In the past few decades, the flat-field spectrometers have been widely used for exploring various intriguing research topics especially in the regime of extreme ultraviolet or soft X-ray, e.g. tokamak plasmas [1], laser-produced warm-dense matters [2], stellar interior properties [3], magnetic confinement fusion problems[4,5], synchrotron radiation light source development [6,7] and so on. The technique is crucial for providing the high spectroscopic resolution in the physical, chemical, photonic and biological sciences and technologies.

Such a spectrometer implements a grating with varied groove density, typically on a concave substrate to achieve a quasi flat-field in the detector plane, and then to deliver high energy resolution through optimizing the coefficients of the variable line spacing (VLS) for the grating. However this type of grazing incidence spectrometer provides the correction to the optical aberrations only in its meridional coordinate but not in the sagittal one, thus still owns significant astigmatism. The meridional rays of the beam are well focused at the detector which is separated from the sagittal focus, displaying a meridionally focused but sagittally diverged 2D spectrograph. So the efforts to achieve better sagittal beam distributions in the detection domain to improve the spectral intensity and acquisition efficiency have never been stopped.

In 1979, G. Tondello demonstrated the stigmatic condition for a spectrometer through the combination of a toroidal mirror and a spherical grating in grazing incidence geometry [8]. In 1992, P. Fan and Z. Zhang replaced the toroidal mirror in the design above with a pair of cylindrical and spherical mirrors, and changed the spherical grating with constant groove density to one with variable line spacing [9]. In 1985, Michael C. Hettrick designed a EUV spectrometer working on a satellite [10]: a pre-focused spherical mirror was utilized to converge the incidence beam beyond a VLS grating forming a virtual source, where the nearly normal incidence geometry is applied, reducing the optical aberration significantly but leading to severe declining of the reflectivity. In 2005, P. Nicolosi, L. Poletto et. al. developed an optical system similating a Kirkpatrick Baez configuration, containing a spherical mirror and a spherical VLS grating placed sequentially and orthogonally, which could provide the flat-field in the focal plane while somehow restrict the astigmatism simultaneously[11]. In 2014, T. Warwick and Y. Chuang et. al. designed a two-dimensional soft X-ray spectrometer implementing Wolter type pre-focusing[12]. In 2015, E. A. Vishnyakov and A.N. Shatokhin employed a normal-incidence multi-layer spherical mirror to replace the gold-coated mirror in Hettrick's design to enhance the reflectivity, and used a better optimized VLS grating to reduce astigmatism, however the bandwidth of the spectrometer is inevitably limited due to the multi-layer coating[13]. In 2016, J. Dvorak, and I. Jarrige et. al. adopted Hettrick-Underwood spectrometer design using an extra plane mirror to fix the beam outgoing direction. The defocus and coma of the spectrometer are well compensated, and its erect focal plane would minimize any necessary mechanical motion of the detector [14].

The "water window" spanning the wavelength range of 2-5nm, could provide the excellent contrast imaging for Carbon (C) or Oxygen (O) atoms and related structures. This outstanding property could be utilized to image and analyze the biological cells or micro-structures in vitro and potentially in vivo. Thus based on all these previous works, we designed a novel flat-field spectrometer optimized in "water window" through systematic investigation of the intrinsic optical nature to exploit its ultimate performance. And the manuscript is organized as the following:

i) The 2nd section introduced the numerical simulation and algorithm to achieve the best meridional focal curve for the spectrometer with various object distances, and then to optimize the sagittal focal curve to approach the meridional one nicely. Especially, the parameters for evaluating the quality of the meridional or sagittal focal curve are well defined and discussed.

ii) The 3rd section explicitly presents the systematic design of the proposed spectrometer using the algorithm in Section 2, to achieve the desired resolving power in the dispersive coordinate while to eliminate astigmatism to improve the spectral intensity.

iii) Finally, we made a more general and summarizing remark regarding to our design, and discussed about the potential research direction and development in the future.

## Ⅱ. Numerical Simulation

As demonstrated in Fig.1 (a), the grating on a concave substrate with VLS groove density is the core of the grazing incidence spectrometer to achieve an excellent "flat field" in its detector plane, while using a plain grating with constant groove density hardly achieves that. The coefficients of the VLS grating ($b_i$) are optimized through elimination of the optical aberrations of various orders in the meridional coordinate, and its groove density is varied and could be expressed as [15]:

$$n(w) = \frac{1}{d_0}\left(1 + \frac{2b_2}{R}w + \frac{3b_3}{R^2}w^2 + \frac{4b_4}{R^3}w^3 + \ldots\right) \quad \textbf{(1).}$$

Where $w$ is the meridional coordinate respected to the center of the grating, $d_0$ is the groove spacing at $w=0$, and $R$ is the meridional radius of the substrate (which is differentiated from the sagittal radius $-\rho$, thus the substrate of the grating is actually in a toroidal profile).

Letting $D_0 = \frac{1}{d_0}, D_1 = \frac{2b_2}{d_0 R}, D_2 = \frac{3b_3}{d_0 R^2}, D_3 = \frac{4b_4}{d_0 R^3}$, Eq. (1) is simplified to:

$$n(w) = D_0 + D_1 w + D_2 w^2 + D_3 w^3 \quad (2).$$

Where $D_0$ is the grooved line density (the reciprocal of $d_0$) at the center of the grating. According to Fermat's principle for geometrical optics, the optimal imaging in meridional coordinate could be achieved through zeroing the first order derivative of the light path function connecting the light source and the image via optics (since the grating is a dispersive optics, so various wavelengths are associated with different preferable optical paths) [16]. And ideally the F-terms (refer to Eq. (21)), especially the first few dominants should satisfy the following equations crossing all over the wavelength range:

$$F_{100} = -\sin\alpha - \sin\beta + D_0 m\lambda = 0 \quad (3),$$

$$F_{200} = \frac{1}{2}\left(\frac{\cos^2\alpha}{r_m} - \frac{\cos\alpha}{R}\right) + \frac{1}{2}\left(\frac{\cos^2\beta}{r'_{20}} - \frac{\cos\beta}{R}\right) - D_1 m\lambda \frac{1}{2} = 0 \quad (4),$$

$$F_{300} = \left(\frac{\cos^2\alpha}{r_m} - \frac{\cos\alpha}{R}\right)\frac{\sin\alpha}{2r_m} + \left(\frac{\cos^2\beta}{r'_{20}} - \frac{\cos\beta}{R}\right)\frac{\sin\beta}{2r'_{20}} - D_2 m\lambda \frac{1}{3} = 0 \quad (5),$$

$$F_{400} = \frac{1}{8}\left[\frac{4\sin^2\alpha}{r_m^2}\left(\frac{\cos^2\alpha}{r_m} - \frac{\cos\alpha}{R}\right) - \frac{1}{r_m}\left(\frac{\cos^2\alpha}{r_m} - \frac{\cos\alpha}{R}\right)^2 + \frac{1}{R^2}\left(\frac{1}{r_m} - \frac{\cos\alpha}{R}\right)\right]$$
$$+ \frac{1}{8}\left[\frac{4\sin^2\beta}{r'^2_{20}}\left(\frac{\cos^2\beta}{r'_{20}} - \frac{\cos\beta}{R}\right) - \frac{1}{r'_{20}}\left(\frac{\cos^2\beta}{r'_{20}} - \frac{\cos\beta}{R}\right)^2 + \frac{1}{R^2}\left(\frac{1}{r'_{20}} - \frac{\cos\beta}{R}\right)\right]$$
$$- D_3 m\lambda \frac{1}{4} = 0$$

$$(6).$$

Where $\alpha$ is the incidence angle; $\beta$ is the diffraction angle; $m$ is the order of diffraction (typically $m=1$ is used in a spectrometer design); $\lambda$ is the wavelength; $r_m$ is the meridional object distance; $r'_{20}$ is the meridional image distance; and $D_i$ are the VLS coefficients defined in Eq. (2). More specifically, the equation of $F_{100}$ is actually the grating formula; the $F_{200}$ one is related to the meridional focus, and could be utilized to characterize the "defocus" over the whole spectral range; $F_{300}$ or $F_{400}$ terms are associated with the "coma" or "spherical aberration" respectively.

### A. Achieve the optimal flat-field in "water window"

While it would be nice if Eq. (4) is satisfied throughout the whole spectral range but not possible, so let it be at the central wavelength. Thus when the meridional object distance $r_m$, beam incident angle $\alpha$, and image distance $r'_{20}(\lambda_0)$ (at the center wavelength $\lambda_0$) are preset, $F_{200}(\lambda_0)=0$ would lead to:

$$D_1(\lambda_0) = \frac{1}{m\lambda_0}\left(\frac{\cos^2\alpha}{r_m} - \frac{\cos\alpha}{R}\right) + \frac{1}{m\lambda_0}\left(\frac{\cos^2\beta(\lambda_0)}{r'_{20}(\lambda_0)} - \frac{\cos\beta(\lambda_0)}{R}\right) \quad (7).$$

Where the first order VLS coefficient $D_1$ is a function of the meridional radius $R$. For the specific values of $R$, $D_1$ being fixed (according to Eq. (7)), the meridional image distances for the entire wavelength range could be calculated then, via re-arrangement of Eq. (4):

$$r'_{20}(\lambda) = \frac{\cos^2\beta(\lambda)}{D_1 m\lambda - \left(\frac{\cos^2\alpha}{r_m} - \frac{\cos\alpha}{R}\right) + \frac{\cos\beta(\lambda)}{R}} \quad (8).$$

Which is wavelength dependent obviously and could be casted into Cartesian coordinates in the principal (i. e. meridional) diffraction plane of the VLS Grating:

$$\begin{aligned} x_{20}(\lambda) &= r'_{20}(\lambda)\sin\beta(\lambda) \\ y_{20}(\lambda) &= r'_{20}(\lambda)\cos\beta(\lambda) \end{aligned} \quad (9).$$

The two-dimensional coordinates: $[x_{20}(\lambda), y_{20}(\lambda)]$ in the principal plane are the theoretically meridional focal spots of various wavelengths, forming an ideal focal curve. The best straight fitting line by using these points would represent the optimal meridional focal line for the detector (i.e. the intersection in-between the meridional plane and the detector), then the distance from the detector (the corresponding impact spots for various wavelengths) to the grating center $-r'_{detector}(\lambda)$ and its orientation in the principal plane could be determined. A mean square root value is introduced to characterize how the realistic detector plane approaches to the ideal focal plane (or curve) by concerning of $N$ different sampling wavelengths:

$$\delta_m = \sqrt{\frac{\sum\left(r'_{20}(\lambda) - r'_{detector}(\lambda)\right)^2}{N}} \quad (10).$$

A smaller value of $\delta_m$ is corresponding to a smaller radial separation in-between the beam colliding spot on the detector and the actual meridional focal spot, indicating a better flat-field condition is achieved i.e. the defocusing within the specific wavelength range is minimized.

Implementing a concrete set of parameters: e.g. $D_0$=24000 ln/cm, $r_m$=1000cm, $r'_{20}(\lambda_0)$=200cm at $\lambda_0 = 3.5$nm and $\alpha$=89.124°, the optimal flat-field searching algorithm could be launched within the water window ($\lambda$=2~5nm). As illustrated by Fig1.(b), various focal curves associated with different values of '$R$' used in the design are plot in a same principal plane which lead to various values of $\delta_m$, and the minimal value is achieved at the optimal meridional radius $R$ of 52729cm (red circle). Generally speaking, each set of the parameters would only lead to a unique optimal meridional radius of $R$ via the above scheme for minimizing the value of $\delta_m$.

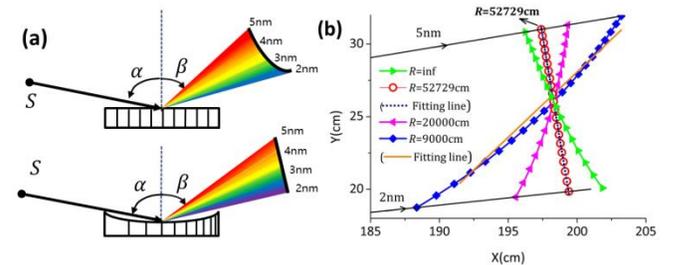

Fig.1 (a) The schematic diagram of a VLS Grating on the concave substrate to achieve an excellent flat-field condition in meridional plane within the spectral range of 2-5nm (the lower one), is compared with a plane grating with constant groove density whose focal spots for the same spectral range lie in a curved line (the upper one). (b) The change of the meridional focal curves associated with different meridional radii $R$ being applied. And for each case, the following parameters used in simulation are identical: $D_0$=24000 ln/cm, $r_m$=1000cm $\alpha$=89.124°, and the image focal length $r'_{20}(\lambda_0)$=200cm at the central wavelength of $\lambda_0 = 3.5$nm. The straight fitting lines represent the best detector plane for each $R$, where the value $\delta_m$ is the magnitude of "defocus" over the whole spectral range (defined in text). The fitting lines for $R$=9000 or 52729cm are depicted, while the diffraction beams of 2nm or 5nm are shown simultaneously. The value $\delta_m$ for various cases are: $\delta_m$=0.01346cm ($R$=inf), $\delta_m$=0.0018cm ($R=52729$cm), $\delta_m$=0.22cm ($R=20000$cm), $\delta_m$=0.6381cm ($R=9000$cm).

The above discussion explicitly explains how to optimize the value of $R$ and coefficient $D_1$ when the object/image distances, beam incident angle, and the grating groove density (at the center) are fixed. Furthermore according to Eq. (5-6), the ideal VLS coefficients of $D_2$ and $D_3$ should be wavelength dependent too, while their optimal values could be obtained at the central wavelength via $F_{300}(\lambda_0)=0$ and $F_{400}(\lambda_0)=0$.

## B. General discussion for various object distances – "$r_m$"

The scheme used to search for the best flat-field condition in meridional coordinate could be extended to more general case, e.g. implementing different object distances - "$r_m$", while the values of the grating groove density $D_0$ and image distance $r_{20}'(\lambda_0)=200$cm are kept the same as those at $r_m=1000$cm. Especially the ideal spectral resolving powers of the spectrometer (for various $r_m$) are pre-assumed to be identical for all of them according to (while the optical aberrations, fabrication or alignment errors in the system were not concerned):

$$A_{ideal} = \frac{\lambda \cdot r_m}{\sigma_s^{[FWHM]} cos\alpha \, d_0} \quad (11).$$

Where $\sigma_s^{[FWHM]}$ is the size of the light source (at the full-width of the half maximum), and the other parameters were previously defined. Eq.(11) could be used to relate the specific meridional object distance $r_m$ to the corresponding incident angle $\alpha$. For example, in order to achieve a resolution of $A\sim12000$ at $\lambda_0=3.5$nm for a source size of $\sigma_s^{[rms]}=200\mu m$ (r.m.s), we have (where $r_m$ is in the unit of 'center-meter'):

$$cos\alpha = \frac{\lambda_0 \, r_m}{2.35\sigma_s^{[rms]} \cdot A \cdot d_0} = 1.49\times10^{-5} r_m \quad (12).$$

Thus the general procedures to achieve the optimal meridional focal curve for a spectrometer are: (i) Identify the source size $\sigma_s^{[rms]}$, object distance $r_m$, image distance $r_{20}'(\lambda_0)$ (at $\lambda_0$), wavelength range, and the groove density of the grating $D_0$; (ii) Specify the incident angle $\alpha$ according to Eq. (12) to achieve the desired spectral resolution at the central wavelength; (iii) Evaluate the defocus $\delta_m$ within the whole spectral range using Eq. (7-10) and find out the minimum, thus the optimal meridional radius $R$ and the associated $D_1$ could be determined simultaneously.

Utilizing the procedures, the meridional focal curve, the optimal fitting line and its orientation in the principal plane could be obtained for different meridional object distance $r_m$. In Fig. 2(a), the best focal curves for various $r_m$ (40m, 20m, 10m, 5m) along with an identical image distance $r_{20}'(\lambda_0)$=200cm are plot all-together at the detector domain. According to Eq. (12), a smaller $r_m$ would be correlated to a smaller grazing incident angle (or bigger incidence angle $\alpha$ in complementary). Fig. 2(a) clearly illustrates that the optimal meridional focal planes for various $r_m$ are associated with different inclination angles in the detector domain, and the change of the inclination angle vs. $r_m$ is demonstrated in Fig.2 (b) further. When $r_m$ increases, the tilt angle of the detector plane experiences a transition from "forward" (the slope of the fitting line is negative) to "backward" (the slope of the fitting line is positive) at the source distance $r_m\sim20$m, and again from "backward" to "forward" at $r_m\sim46$m (refer to the two thin vertical line segments in Fig.2 (b)).

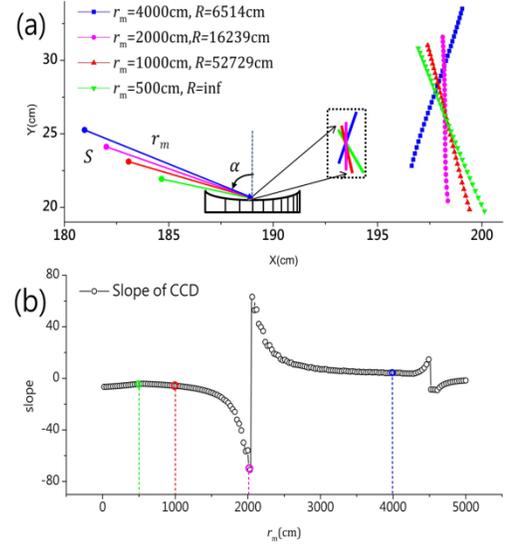

Fig.2 (a) The best meridional focal curves (in various color legends) for different meridional object distances $r_m$ (associated with the various incidence angles to maintain the ideal spectral resolution) are achieved using the scheme discussed in text, where the image distance $r_{20}'(\lambda_0)$=200cm is fixed for various $r_m$. (b) The best fitting lines for different $r_m$ could be identified and regarded as the actual detector plane, whose slope in the meridional plane of the grating is plot against $r_m$ reflecting the orientation of the detector in space, the detector plane is tilted towards the projection of the diffraction beam for smaller $r_m$ (e.g. $r_m$=500cm, 1000cm, where the slope of the fitting line is negative) and tilted away from the diffraction beam for larger $r_m$ (e.g. $r_m$=4000cm, where the slope is positive).

## C. Optimization of the sagittal focal curve

We discussed about the scheme to achieve the optimal focal curve in meridional coordinate in previous sections, now switch to the sagittal coordinate concerning the same wavelength range:

$$F_{020} = \frac{1}{r_s} + \frac{1}{r_{02}'} - \frac{cos\alpha + cos\beta(\lambda)}{\rho} = 0 \quad (13).$$

Where $r_s$ or $r_{02}'$ are the sagittal object or image distances respectively, $\rho$ is the sagittal radius of the grating, and it would lead to:

$$r_{02}'(\lambda) = \frac{1}{\frac{cos\alpha + cos\beta(\lambda)}{\rho} - \frac{1}{r_s}} \quad (14).$$

The sagittal focal curve could be converted into Cartesian coordinates as well:

$$\begin{aligned} x_{02}(\lambda) &= r_{02}'(\lambda)sin\beta(\lambda) \\ y_{02}(\lambda) &= r_{02}'(\lambda)cos\beta(\lambda) \end{aligned} \quad (15).$$

Then a parameter $\delta_s$ similar to $\delta_m$ is used to represent the defocus in the sagittal coordinate of the spectrometer:

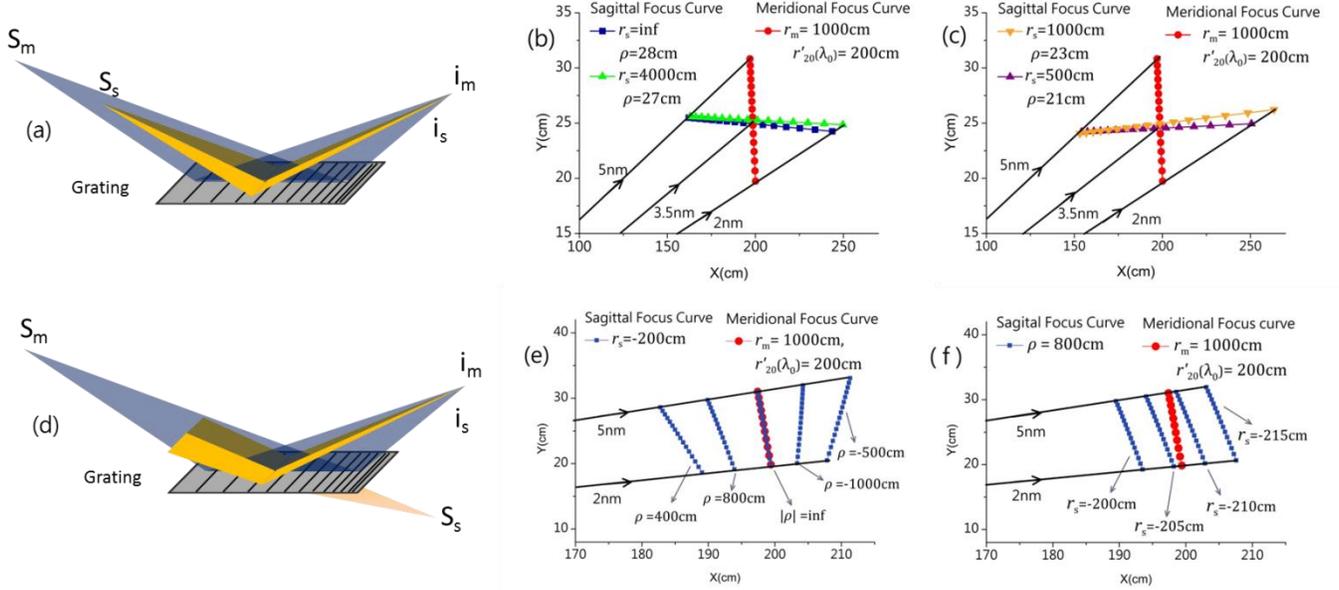

Fig.3 (a) The schematic diagram illustrates the light source with different object distances in its meridional (grey) or sagittal (yellow) coordinates achieves the identical image focal length in these two directions: (a) Both the meridional and sagittal object distances are real (i.e. $r_m > 0$ and $r_s > 0$); (d) The meridional object distance is real, while the sagittal one is virtual (i.e. $r_m > 0$ and $r_s < 0$). The optimal meridional focal curve (red-spot) is presented in each diagram (b, c, e, f) as the control group, where $r_m$=1000cm, $r_{20}'(\lambda_0)$=200cm (at central wavelength), and the other parameters are the same as those for the optimal case in Fig.1 ($R$ = 52729cm). And the sagittal focal curves for various cases are presented in specific plots for comparison. (b) The sagittal focal curves for $r_s=\inf$, $\rho$=28cm and $r_s$=40m, $\rho$=27cm. (c) The sagittal focal curve for $r_s$=10m, $\rho$=23cm and $r_s$=5m, $\rho$=21cm. (e) The sagittal focus curves for fixing $r_s = -200$cm while changing $\rho$. (f) The sagittal focus curves for fixing $\rho$=800cm while changing $r_s$.

$$\delta_s = \sqrt{\frac{\sum (r_{02}'(\lambda) - r_{detector}'(\lambda))^2}{N}} \quad (16).$$

The smaller is $\delta_s$, the closer the sagittal focal curve approaches to the plane of the detector, as well to the meridian focal curve. Thus through this scheme, the astigmatism of the optical system could be well eliminated.

According to Eq. (14), the magnitude of $r_{02}'(\lambda)$ is relevant to both the source distance in sagittal coordinate $r_s$ and the sagittal radius of the grating $\rho$. Here we would like to consider the spatial separation in-between the sagittal source point and the meridional one to make the discussion more general, where $r_s > 0$ is related to the real sagittal object distance (refer to Fig. 3(a-c)) and $r_s < 0$ is the virtual case (Fig. 3(d-f)). For all cases in Fig.3, the meridional focal curve is identical and optimized at $r_m$=1000cm and $r_{20}'(\lambda_0)$=200cm, and plot in each sub-figure as the reference signal (red dots in Fig. 3 (b, c, e, f)).

Fig.3 (a) presents the schematic diagram for the real sagittal source point ($r_s > 0$), where the source distances in the meridional or sagittal coordinates could be different. For different sagittal source distances $r_s$, the sagittal radius $\rho$ is optimized to cast the specific sagittal focal curves at the detector. Obviously they would intersect with the meridional focal curve at the center wavelength only; while at the other wavelengths, the spatial separations in-between the sagittal image focal points and the corresponding meridional image focus are still pretty large, indicating that the astigmatism within the spectral range is not well eliminated i.e. the sagittal image focal curve is far from being optimized (refer to Fig. 3(b-c)).

While for the virtual sagittal source point case ($r_s < 0$), the sagittal rays of the beam would converge and achieve the beam waist behind the grating (refer to Fig.3 (d), the meridional parameters are the same as Fig.3 (a)). Fig.3 (e-f) demonstrated how the sagittal object distance $r_s$ and radius $\rho$ would influence the sagittal focal curves. In Fig.3 (e), $r_s = -200$cm is kept as a constant while the value of $\rho$ changes, it is observed that both the position and tilt angle of the sagittal focal curve change associatively. Especially when $\rho$ becomes infinity representing a tangentially cylindrical grating, the sagittal focal curve would become a circle with radius $|r_s|$ surrounding the center of the grating (refer to Eq. (14)). Fig.3 (f) shows the case that $r_s$ is varied from -200cm to -215cm, while $\rho$=800cm is a constant. The sagittal focal curves are observed to move further away from the grating gradually, however their inclination angles change very little. So according to the above investigation, we find out that $\rho$ affects both the position and the inclination angle of the sagittal image focal curve, while $r_s$ mainly influences the position. Therefore the combination of any arbitrary value of $\rho$ or $r_s$ would lead to various shapes and locations of the sagittal focal curve, and the optimal one could always be achieved through this searching algorithm.

Implementing the scheme described in Fig. 3(d-f), the optimal sagittal focal curves could be identified for various meridional object distances of $r_m$ (e.g. 4000cm, 2000cm, 1000cm and 500cm). In Fig. 4, the optimal meridional and sagittal focal curves for these four different $r_m$ along with identical $r_{20}'(\lambda_0)$=200cm are demonstrated overlapping pretty well, where two non-optimized sagittal focal curves are included in each plot for comparison. The key parameters for each of them are highlighted, while more explicit parameters and the value of "quality assessment" (i.e. $\delta_m$ or $\delta_s$) for various cases are presented in Table.1 and in the next section for further discussion.

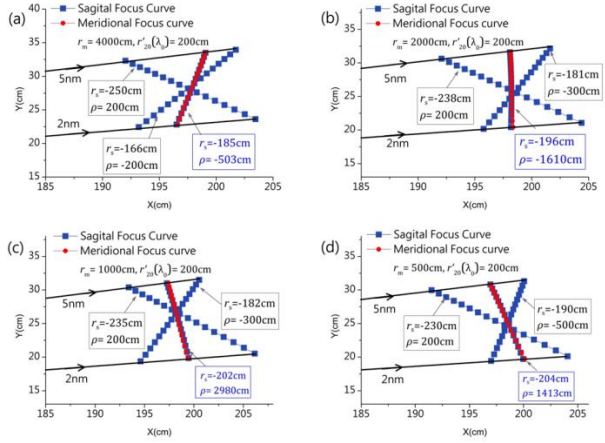

Fig.4 The sagittal focal curves (blue square) are plot together with the optimal meridional focal line (red disk) for different configurations (mainly associated with different $r_m$, and the explicit list of the parameters are outlined in Table 1). For each case, the optimal sagittal focus curve well overlaps with the meridional one, while two non-optimized sagittal focal curves are presented in the same plot for comparison. (a) $r_m$=4000cm associated with the optimal sagittal parameters: $r_s$=-185cm, $\rho$=-503cm. (b) $r_m$=2000cm associated with the optimal sagittal parameters: $r_s$=-196cm, $\rho$=-1610cm. (c) $r_m$=1000cm with $r_s$=-202cm, $\rho$=2980cm. (d) $r_m$=500cm with $r_s$=-204cm, $\rho$=1413cm.

## III. System design

According to the discussion in the previous section, we would like to design a delicate spectrometer operating in "water window", which could not only achieve a decent flat-field in meridional coordinate to deliver high spectral resolution, but also eliminate the astigmatism in sagittal coordinate to enhance the detection efficiency and spectral intensity. The flat-field is achieved in meridional coordinate via a variable line-spacing grating on a toroidal substrate; while the sagittal object distance $r_s < 0$ indicates that a virtual light source is preferable in sagittal direction, which could be realized by using a pre-focusing cylindrical mirror in front of the grating.

The proposed design of a realistic spectrometer possessing the aforementioned merits is illustrated in Fig.5 (a): a vertically placed cylindrical mirror is combined with a horizontal placed VLS toroidal grating, with an appropriate spatial separation in-between them. The in-coming beam is focused by the cylindrical mirror horizontally, while propagates down to the grating as a free beam vertically, with mild divergence. With respect to the grating, the source points separate in its horizontal or vertical coordinates: the vertical one is located within the meridional (or dispersive) coordinate, at the far field; while the horizontal one forms a virtual sagittal source of a converging beam beyond the grating within its non-dispersive coordinate.

In order to achieve this, the radius of the cylindrical mirror should satisfy the following equation:

$$R_{cm} = \frac{2}{\cos\alpha_c} / \left( \frac{1}{r_m - d} + \frac{1}{d + |r_s|} \right) \quad (17).$$

Where $r_m - d$ or $d + |r_s|$ are the effective source or image distances for the cylindrical mirror in its meridional coordinate (where $r_m > d > 0$ and $r_s < 0$ per previous discussion), $d$ is the distance between the cylindrical mirror and the grating, $\alpha_c$ is the incident angle for the cylindrical mirror (which is set equal to the incidence angle of the grating for convenience, i.e. $\alpha_c = \alpha$).

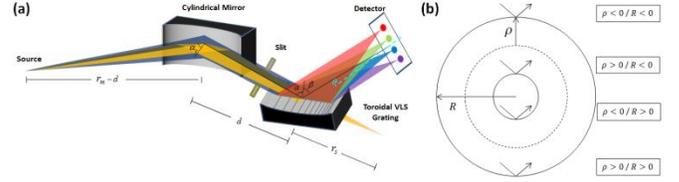

Fig.5 (a) Schematic layout of the system design for the sagittal-enhanced flat-field spectrometer in "water-window". (b) Options for the toroidal substrate profile of the grating.

In design of such a spectrometer, we set the spectral resolution above ~12000 at the central wavelength of $\lambda_0$=3.5nm, and optimize the design throughout the "water window" spectrum by considering the light source size of 200$\mu$m (r.m.s), and divergence angle of 20μrad (r.m.s). The concrete design parameters of the spectrometer for the object distances ($r_m$) of 40m, 20m, 10m and 5m are listed in Table.1. Especially for the substrate of the grating, two over four available toroidal surfaces were adopted for the design: $\rho<0$, $R>0$ (column A、B in Table.1) or $\rho>0$, $R>0$ (C、D in Table.1), associated with "convex (sagittal) – concave (meridional)" or "concave (sagittal) – concave (meridional)" surface profiles respectively (refer to Fig.5 (b)).

**Table1. The design parameters of the optimized spectrometer for four different source distances (the corresponding schematic layout is presented in Fig. 5).**

| Configuration | A | B | C | D |
|---|---|---|---|---|
| Cylindrical Mirror | | | | |
| $r_m - d$ (cm) | 3960 | 1960 | 960 | 460 |
| $d$ (cm) | 40 | 40 | 40 | 40 |
| $R_{cm}$ (cm) | 6964 | 13779 | 25287 | 41718 |
| $\alpha$ ($\alpha_c$) | 86.494° | 88.248° | 89.124° | 89.560° |
| Toroidal VLS Grating | | | | |
| $r_m$ (cm) | 4000 | 2000 | 1000 | 500 |
| $r_s$ (cm) | -185 | -196 | -202 | -204 |
| $r_{20}'(\lambda_0)$ (cm) | 200 | 200 | 200 | 200 |
| $R$ (cm) | 6514 | 16239 | 52729 | inf |
| $\rho$ (cm) | -503 | -1594 | 2980 | 1413 |
| $D_0$ (ln/cm) | 24000 | 24000 | 24000 | 24000 |
| $D_1$ (ln/cm$^2$) | 205.1 | 224.8 | 235.1 | 240.15 |
| $D_2$ (ln/cm$^3$) | 1.693 | 1.698 | 1.749 | 1.748 |
| $D_3$ (ln/cm$^4$) | 0.017 | 0.011 | 0.011 | 0.012 |
| Footprint [FWHM] on the Grating Surface | | | | |
| $w$ (cm) | 3.158 | 3.499 | 4.350 | 6.711 |
| $l$ (cm) | 0.1619 | 0.0858 | 0.0514 | 0.0383 |
| Slope Errors | | | | |
| $SE_m$ (μrad) | 0.5 | 0.5 | 0.5 | 0.5 |
| $SE_s$ (μrad) | 2 | 2 | 2 | 2 |
| Quality Assessment | | | | |
| $\delta_m$ (cm) | 0.0033 | 0.0031 | 0.0018 | 0.0029 |
| $\delta_s$ (cm) | 0.0302 | 0.0416 | 0.0749 | 0.0484 |

In order to calculate the resolving power of the spectrometer, we first evaluate the line width of the diffraction beam distributed at the detector:

$$\sigma_d^{[FWHM]} = \sigma_S^{[FWHM]} \frac{\cos\alpha}{\cos\beta} \frac{r_{20}'}{r_m} \frac{m}{\cos\theta} \quad (18).$$

Where $\theta$ is the defined as the angle in-between the central diffraction beam and the normal of the X-ray detector (Fig.5). And it needs to point that, a reasonable image-to-object magnification should be implemented to ensure that the line width is greater than the pixel size of the detector to guarantee the resolution is realistic.

Then according to differentiation of the grating formula in Eq. (3), the spectral line width could be expressed as the image line distribution at the detector:

$$\Delta\lambda_d^{[FWHM]} = \frac{d_0 \cos\beta}{m}\Delta\beta = \frac{d_0 \cos\beta}{m}\frac{\sigma_d^{[FWHM]}\cdot\cos\theta}{r_{20}{'}} \quad (19).$$

Implementing Eq. (18) to Eq. (19), the spectral line width due to the light source size could be calculated:

$$\Delta\lambda_{so} = \Delta\lambda_d^{[FWHM]} = \frac{d_0 \cdot \sigma_S^{[FWHM]}\cdot\cos\alpha}{r_m} \quad (20).$$

Eq. (18-20) actually shows how Eq. (11) is derived (since $A_{ideal} = \lambda/\Delta\lambda_{so}$), indicating that a Gaussian distribution beam in an aberration-free optical system could achieve the ideal spectral resolution which is only limited by its source size. However, in a realistic optical system, the optical aberrations are non-negligible, which would broaden the line spread width of an ideal Gaussian distribution beam substantially. The aberration broadening effect in the meridional coordinate (dispersion direction) could be expressed as:

$$\Delta y_{ijk} = \frac{r_{20}{'}}{\cos\beta}\frac{\partial}{\partial w}\left[F_{ijk}w^i l^j\right] \quad (21).$$

Where $w$ is the illuminated meridional length of the grating, $l$ is the illuminated sagittal length, and $F_{ijk}$ defines the optical aberrations in various orders (the subscript - '$i$' or '$j$' denotes the meridional or sagittal coordinate respectively). Then the spectral distribution broadening due to the aberration in the system could be evaluated, via combining Eq. (19) with Eq. (21):

$$\Delta\lambda_{ijk} = \frac{d_0 \cos\beta}{m\cdot r_{20}{'}}\cos\theta\left(\frac{\Delta y_{ijk}}{\cos\theta}\right)$$
$$= \frac{d_0 \cos\beta}{m r_{20}{'}}\frac{r_{20}{'}}{\cos\beta}\frac{\partial}{\partial w}\left[F_{ijk}w^i l^j\right] = \frac{d_0}{m}\frac{\partial}{\partial w}\left[F_{ijk}w^i l^j\right] \quad (22).$$

And the dominant meridional aberration terms are ($l=0$):

$$\Delta\lambda_{200} = \frac{d_0}{m}2wF_{200} \quad (23),$$

$$\Delta\lambda_{300} = \frac{d_0}{m}3w^2 F_{300} \quad (24),$$

$$\Delta\lambda_{400} = \frac{d_0}{m}4w^3 F_{400} \quad (25).$$

The explicit expressions of $F_{200}$, $F_{300}$ and $F_{400}$ were already given in Eq. (4-6), which are independent of $w$ and $l$. The optical aberrations are set to zero at the central wavelength, i.e. $F_{ijk}(\lambda_0) = 0$, while the defocus $F_{200}$ has been minimized to achieve an optimal flat field (i.e. $\delta_m$ is minimized), the coma $F_{300}$ and the spherical aberration $F_{400}$ are eliminated to as small as possible for the whole "water window" by employing the scheme discussed in section 2(A).

Moreover, the optical fabrication error (including the slope error and surface roughness etc.) should be taken into account, which broadens the spectral line width by [17]:

$$\Delta\lambda_{se} = 2.355\cdot SE_E \cdot \frac{d_0}{m}\left(\cos\alpha + \cos\beta\right) \quad (26).$$

Where $SE_E$ represents the meridional slope error of the grating, while the surface roughness has little impact to the spectral distribution, but would strongly influence the reflectivity of the beam at the surface.

The ideal spectral resolution for an aberration-free Gaussian beam is given by Eq. (11) previously. When the overall systematic errors in the spectrometer are inclusive, the resolution could be evaluated [18]:

$$A_{theory} = \frac{\lambda}{\Delta\lambda_{sum}} \cong \frac{\lambda}{\sqrt{\Delta\lambda_{so}^2 + \Delta\lambda_{se}^2 + \left(\Delta\lambda_{200} + \Delta\lambda_{300} + \Delta\lambda_{400}\right)^2}} \quad (27).$$

The four concrete spectrometer models in Table.1 (A-D) could be used to calculate the various spectral distribution terms via implementing Eq. (20, 23-25, 26), and the results are shown in Fig. 6 (a-d). Apparently the source size term is overwhelming and appearing as a constant within the spectral range (since the source size is assumed to be constant throughout the spectral range). The slope error term is the second largest component and more or less constant as well. Among the three primary optical aberration terms, the defocus $\Delta\lambda_{200}$ is well confined indicating the excellent flat-field condition is achieved; the spherical aberration $\Delta\lambda_{400}$ is also pretty small, fluctuating around the zero-crossing and negligible; the value of coma $\Delta\lambda_{300}$ is relatively larger than the other two ($\Delta\lambda_{200}$ or $\Delta\lambda_{400}$) for the configuration (A) in Table.1 (Fig.6(a)), and decreases substantially for the configuration (B, C, D) when the magnification increases (Fig.6(b-d)), i.e. $r_m$ is decreased since $r_{20}{'}$ is constant for all cases. Meanwhile the corresponding resolving powers for the configurations in Fig. 6(a-d) are calculated and exhibited in Fig. 6(e-h) respectively, where in each diagram the ideal spectral resolution $A_{ideal} = \lambda/\Delta\lambda_{so}$, the theoretical resolution $A_{Theory} = \lambda/\Delta\lambda_{sum}$, and the result from the ray-tracing program $A_{Trace}$ (Fig. 7) are overlaid for comparison.

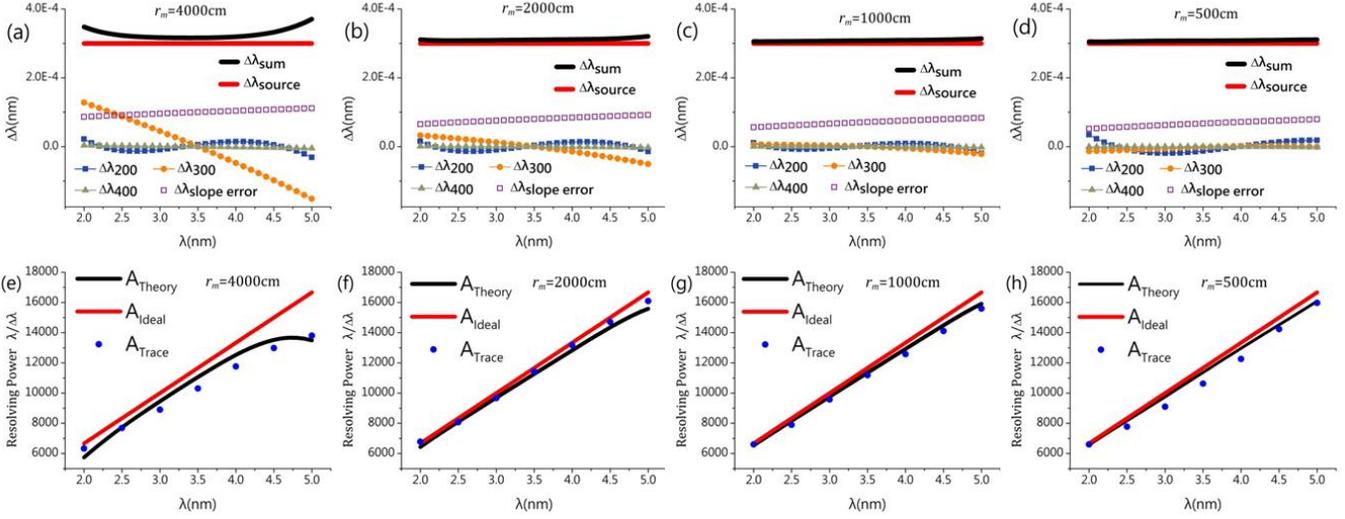

Fig.6. The simulated results of the major factors which influence the resolving power of the spectrometer, including the source size (thick red), the optical fabrication error (empty square), the optical aberrations – defocus (filled square), coma (filled circle), spherical aberration (grey triangle) and the overall (thick black). The results for various object distances are presented, (a) $r_m=4000$cm , (b) $r_m=2000$cm , (c) $r_m=1000$cm , and (d) $r_m=500$cm ; and the image distance for all four cases is identical: $r_{20}'(\lambda_0)=200$cm . The corresponding resolving powers of (a-d) are calculated and presented in (e-h) respectively, where for each case three types of the spectral resolutions are shown: $A_{Ideal}=\lambda/\Delta\lambda_{so}$ , $A_{Theory}=\lambda/\Delta\lambda_{sum}$ and $A_{Trace}$ (obtained from the ray-tracing program, e.g. Fig. 7), for (e) $r_m=4000$cm , (f) $r_m=2000$cm , (g) $r_m=1000$cm , and (h) $r_m=500$cm .

Additionally, the ray-tracing program for the geometric optics – 'Shadow' is utilized to demonstrate the spectral resolution at 2nm, 3nm, 4nm, 5nm respectively for the configuration (C) in Table.1. The bottom part of Fig. 7 shows the spectral distributions at the optimal detector plane for the whole "water-window" (2-5nm), where the length scales in the meridional (larger) and sagittal (smaller) directions are different ($140\text{mm}\times0.8\text{mm}$). It is worthwhile to point out that sagittal distribution profiles of the four wavelengths are approximately uniform, indicating the astigmatism of the spectrometer is well restricted, otherwise for un-compensated astigmatism case, the sagittal focal size at various wavelengths would be very different. Fig.7 (a-d) exhibit the spectral distribution and resolution for each wavelength individually (2, 3, 4, 5 nm in terms of $\lambda$ and $\lambda+\Delta\lambda$), which are traced in a $400\mu\text{m}\times400\mu\text{m}$ square detector domain. The FWHM widths in both directions are illustrated in the plots, especially the meridional ones could be used to evaluate the realistic spectral resolution $A_{Trace}=\lambda/\Delta\lambda$ for each wavelength (the results presented in Fig. 6).

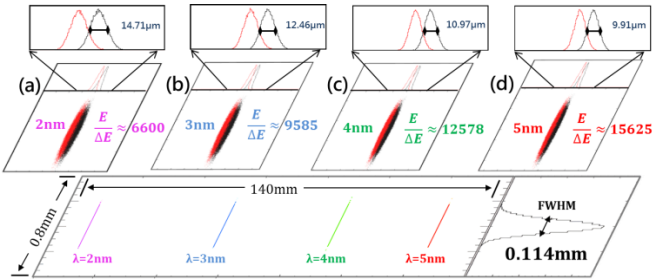

Fig.7. The ray-tracing results for the spectrometer configuration (C) in Table.1. The spectral profile distributions at the optimal detector plane for the full wavelength range (2-5nm) are demonstrated at the lower part of the figure, where the detector dimensions are 140mm (meridional) ×0.8mm (sagittal). The ray-tracing results for each individual wavelength of 2nm, 3nm, 4nm and 5nm are presented in (a) to (d), where an identical "domain of interest" is applied at the detector plane for each of them: the meridional size of the spectrograph is 10-15μm (FWHM), the sagittal size is about 110μm (FWHM). Then the resolution power at various wavelengths could be obtained through $A_{Trace}=\lambda/\Delta\lambda$: (a) 6600 at 2nm, (b) 9500 at 3nm, (c) 12500 at 4nm, and (d) 15000 at 5nm.

## IV. Discussion

In summary, we report a novel spectrometer design in combination with a sagittal pre-focusing cylindrical mirror and a toroidal VLS grating, which could not only provide an excellent flat field in the meridional coordinate to achieve the desired resolving power for the whole spectral range, but well eliminate the astigmatism in the sagittal coordinate to enhance the spectral intensity. Our main findings in the current research are: 1) While various meridional object distances $r_m$ are employed in the spectrometer design, the specific incident angle $\alpha$ could be determined to correlate to each value of $r_m$ to maintain a constant spectral resolution. 2) For each $r_m$, there is only one unique set of meridional radius $R$ along with "defocus" correction coefficient $D_1$ (the lowest order of the VLS coefficients) for the grating, which could achieve the optimal meridional focal curve $r_{20}'(\lambda)$ throughout the spectral range; in the meantime "coma" and "spherical aberration" of the system could be well eliminated by optimizing the VLS coefficients $D_2$ and $D_3$. 3) Then the best sagittal focal curve (which approaches the meridional one pretty well) could be achieved, through optimizing the radius of a sagittal pre-focusing cylindrical mirror $R_{cm}$ and the sagittal radius of the grating $\rho$.

Thus the intrinsic optical aberrations presence in a grazing incidence X-ray spectrometer could be well compensated by the meridional VLS coefficients of the grating, and by the astigmatism-elimination scheme described above. Especially, the idea of separating the light source points in its meridional or sagittal coordinates provides high degree of freedom for selecting and optimizing the parameters of the spectrometer, via a rather simple simulation algorithm. Here we demonstrated the scheme by employing a spectrometer design in "water window" with a large source distance (i.e. the image-to-object magnification is less than 1), which is well applicable to a light source split from the beam line of a synchrotron radiation or free electron laser. While the scheme is not limited to this but could be implemented to the design of a compact spectrometer as well. Furthermore, we have the flexibility to pursue an even higher resolving power, by implementing a meridional confinement slit in the incident beam line to achieve a smaller effective source size for the spectrometer (refer to Fig.5 (a)). However, the smaller source size would correspond to a smaller imaging line width at the detector, which needs to be assessed in advance to make sure it greater than the detector's pixel size to guarantee the spectral resolution. More details regarding to this could be found in the supplemental materials.

Although we mainly discussed about the spectrometer design in "water window", the design algorithm actually owns a universal adaptability, which could be easily extended in much broader photon-energy (or wavelength) range through an appropriate modification to the design parameters. And it is also feasible to utilize the scheme to develop a high performance grating monochromator simply by putting a fine silt right across the focal curve of the diffraction beam.

**Funding**. National Science Foundation of China (NSFC) (11475249), and Youth 1000-Talent Program in China (Y326021061).

**Acknowledgment**. The authors thank for the staff and facility support from the Department of Free Electron Laser Science & Technology, Shanghai Institute of Applied Physics, Chinese Academy of Sciences.

See Supplement 1 for supporting content.

## Supplement 1

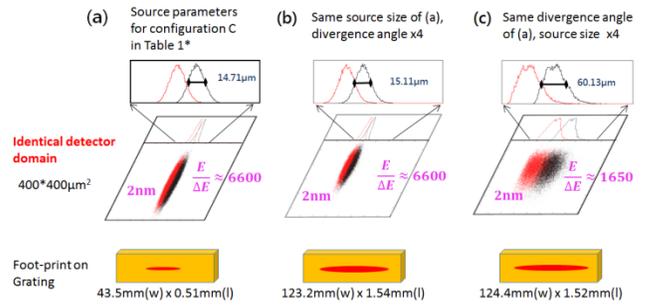

Fig. S1. Additional ray-tracing results for the spectrometer configuration (C) in Table.1 of the main text: (a) using the identical parameters of the configuration (C) in Table .1, where light source size of 200μm (r.m.s), and divergence angle of 20μrad (r.m.s), (b) the same source size while the divergence angle increases 4 times, (c) the same divergence angle while the source size increases 4 times. For each case, the corresponding beam foot-print on the surface of the grating is illustrated underneath, where w or l represents the Meridional or Sagittal coordinate respectively, and all the values are given in the FWHM of the beam.

According to the previous discussion, the light source size of 200μm (r.m.s), and divergence angle of 20μrad (r.m.s) are adopted for the spectrometer design. While the parameters of the configuration (C) in Table 1 (of the main text) are implemented, the spectral resolution of ~6600 could be achieved at λ=2nm, and the beam foot-print on the grating is 43.5mm(w) x 0.51mm(l), where w or l represents the meridional or sagittal coordinate of the grating respectively (Fig.S1(a)). If the beam size is kept the same while the divergence angle increases 4 times, i.e. to 80μrad (r.m.s), the ray-tracing indicates that the foot-print on the grating increases 3 times approximately while the spectral resolution still remains as its original value (Fig.S1(b)). However if the beam divergence is kept the same while the source size increases 4 times, i.e. to 800μm (r.m.s), the ray-tracing exhibits that the resolving power decreases substantially down to ~1650 (Fig.S1(c)). The comparison here demonstrates that the optical aberrations in the spectrometer are well compensated and corrected, so increasing the divergence angle wouldn't influence the resolving power, while only debase the quality of the spectrograph a bit. The size of the light source is the dominant factor restricting the resolution power, which could be further improved via narrowing down the source size, especially the meridional one.

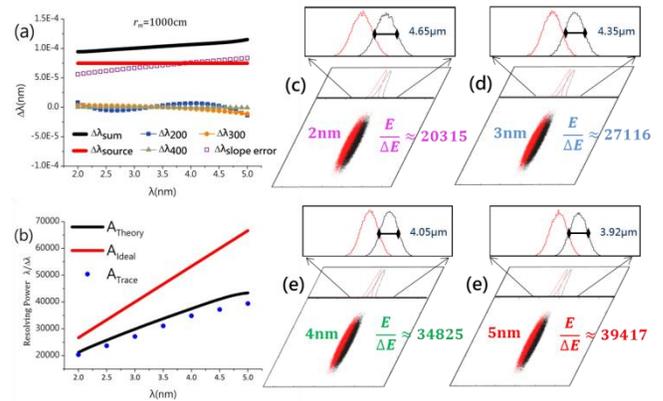

Fig. S2. The ray-tracing results for the spectrometer configuration (C) in Table.1, but with a much smaller meridional source size of 50μm (compared to Fig. 7).

Furthermore as illustrated in Fig. 5(a) of the main text, if a confinement slit is inserted into the incident beam path to narrow down the meridional beam size, which simulates to decrease its effective source size, the spectral resolution would apparently be enhanced further. While the source size(200μm) is replace to 50μm, the various spectral distribution terms ('Footprint' and the items below) along with the spectral resolving powers in Table.1(C) could be recalculated by using the same scheme discussed in Section 3. As demonstrated in Fig. S2, the theoretical spectral resolution ($A_{Theory}$ or $A_{Trace}$) could be improved substantially to 20000-40000 for the "water window" (2-5nm), however it deviates more from the ideal resolution ($A_{Ideal}$) compared to the previous case (Fig. 6). It is mainly associated with the relatively stronger influence due to the optical fabrication slope errors for a much smaller source size used in the spectrometer design. Neverless this demonstrates that, upon scarifying certain amount of beam flux, the spectrometer has potential to further increase the resolution power throughout the spectral range.